\documentclass[aps,pra,reprint,superscriptaddress]{revtex4-2}
\usepackage{amsmath}
\usepackage{amssymb}
\usepackage{amsthm}
\usepackage{bbm}
\usepackage{eucal}
\usepackage{graphicx}[floatfix]
\usepackage{hyperref}
\usepackage{physics}
\usepackage{times}

\RequirePackage{color}

\newcommand{\orcid}[1]{\href{https://orcid.org/#1}{\includegraphics[width=7pt]{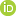}}}

\DeclareMathOperator{\diag}{diag}

\hypersetup{
  colorlinks=true,linkcolor=blue,citecolor=blue,
  filecolor=blue,urlcolor=blue,breaklinks=true
}

\begin{document}

\title{Average scattering entropy of quantum graphs}

\author{Alison A. Silva\orcid{0000-0003-3552-8780}}
\email{alisonantunessilva@gmail.com}
\affiliation{
  Programa de P\'os-Gradua\c{c}\~{a}o Ci\^{e}ncias/F\'{i}sica,
  Universidade Estadual de Ponta Grossa,
  84030-900 Ponta Grossa, Paran\'a, Brazil
}

\author{Fabiano M. Andrade\orcid{0000-0001-5383-6168}}
\email{fmandrade@uepg.br}
\affiliation{
  Programa de P\'os-Gradua\c{c}\~{a}o Ci\^{e}ncias/F\'{i}sica,
  Universidade Estadual de Ponta Grossa,
  84030-900 Ponta Grossa, Paran\'a, Brazil
}
\affiliation{
  Departamento de Matem\'{a}tica e Estat\'{i}stica,
  Universidade Estadual de Ponta Grossa,
  84030-900 Ponta Grossa, Paran\'{a}, Brazil
}

\author{Dionisio Bazeia\orcid{0000-0003-1335-3705}}
\email{bazeia@fisica.ufpb.br}
\affiliation{
  Departamento de F\'{i}sica,
  Universidade Federal da Para\'{i}ba,
  58051-900 Jo\~{a}o Pessoa, Para\'{i}ba, Brazil
}

\date{\today}

\begin{abstract}
The scattering amplitude in simple quantum graphs is a well-known
process which may be highly complex.
In this work, motivated by the Shannon entropy, we propose a methodology
that associates a graph with a scattering entropy, which we call the
average scattering entropy.
It is defined by taking into account the period of the scattering
amplitude which we calculate using the Green's function procedure.
We first describe the methodology on general grounds, and then exemplify
our findings considering several distinct groups of graphs.
We go on and investigate other possibilities, one that contains groups
of graphs with the same number of vertices, with the same degree, and
the same number of edges, with the same length, but with distinct
topologies and with different entropies.
Another possibility we investigate contains graphs of the fishbone type,
where the scattering entropy depends on the boundary conditions on the
vertices of degree $1$, with the corresponding values decreasing and
saturating very rapidly, as we increase the number of elementary
structures in the graphs.\\
\newline
DOI: \href{https://doi.org/10.1103/PhysRevA.103.062208}
{10.1103/PhysRevA.103.062208}
\end{abstract}

\maketitle

\section{Introduction}

The purpose of the present work is to suggest procedure that allows one
to add global information directly related to the complexity of
the scattering on a quantum graph.
The point of view we assume here is inspired by the Shannon
entropy \cite{Book.Shannon.1963}, which can be described as follows: if
we are dealing with a system with $n$ independent accessible states,
with the probability distribution $p_i\, (i=1,2,...,n)$ to access the $i$-th
state, then the Shannon entropy has the form
\begin{equation}
  H_n=\sum_{i=1}^{n} p_i \log_2 \frac{1}{p_i}
  =-\sum_{i=1}^{n} p_i \log_2 p_i,
\end{equation}
with $0 \log_2 0$ defined to be $0$.
In the simple case of a fair die with the six distinct faces painted
with six distinct colors, the Shannon entropy gives $H_6=\log_2 6$,
since the system has six distinct accessible states, each one with
probability $1/6$.
However, if one takes another fair die and paints the six faces with
three distinct colors (for instance, two red, two green and two blue) we
get $H_{3}=\log_2 3$, and if we paint three faces with the color red and
the other three with the color blue, we get $H_{2}=\log_2 2=1$.
This last case is similar to the case of a fair coin, which has only two
distinct accessible states.
With this simple illustration we see that the greater the number of
independent accessible states is, the larger the Shannon entropy is.
In this sense, as one knows, the Shannon entropy provides an interesting
way to measure the informational contents of the system.

Let us now turn our attention to the case of a biased coin, for instance.
Suppose we construct two distinct coins, one with probability $6/10$ to
get heads and probability $4/10$ to get tails, and the other with
probability $7/10$ to get heads, and probability $3/10$ to get tails.
In these two cases, the Shannon entropy will be $0.67301$ and $0.61086$,
respectively.
These values are below the ideal coin and so are poorer than it, but in
the second case, the system is poorer than in the first case, as
expected.
Also, if the coin is $100\%$ biased, such that it always gives heads,
the associated Shannon entropy will vanish, and the system will be the
poorest possible, since it has no informational contents anymore.
If we go further on and consider a continuously biased coin, a coin with
probability $p\in[0,1]$ to get heads, and $1-p$ to get tails, the
associated Shannon entropy will depend on $p$, giving
$H(p)=-p\log_2(p)-(1-p)\log_2(1-p)$.
This entropy vanishes at both $p=0$ and $1$, as expected, and gets the
maximum value at $p=1/2$, showing that the fair coin is the best coin,
at least in the sense of providing information.

The above reasonings are well-known, and the Shannon entropy has been
used in  applications in distinct situations with interesting results.
We recall, for instance, the case where the Shannon entropy works to
compensate the diminishing in the thermodynamics entropy that arises
from the flow of heat against a thermal gradient
\cite{PRL.111.030602.2013} and several other issues devoted to the
inclusion of information into thermodynamics \cite{NP.11.131.2015}.
For instance, in Ref. \cite{PRL.117.240502.2016} the authors employed
information about microscopic degrees of freedom to compensate for
entropy production or work extraction in a quantum system.
We also recall the description of the configurational entropy which was
studied in Ref. \cite{PLB.713.304.2012} and further considered in
different contexts in Refs.
\cite{PLB.737.388.2014,PLB.737.388.2014,PLB.763.434.2016,
PRD.94.083509.2016,PLB.776.78.2018,JMMM.475.734.2019,
PRD.101.105106.2020}, to quote some recent investigations.
We notice, in particular, the use of the configurational entropy to
study very different issues such as the investigation to calculate the
stability of the graviton Bose-Einstein condensate in the braneworld
scenario \cite{PLB.763.434.2016}, the use as a bounding for Gauss-Bonnet
braneworld models \cite{PRD.94.083509.2016}, the internal disposition of
magnetization in magnetic skyrmions \cite{JMMM.475.734.2019}, and the
stability of mesons made of heavy quark/antiquark pairs within the
context of a holographic anti-de Sitter/QCD model for heavy vector
mesons \cite{PRD.101.105106.2020}.
In the present work we focus on another issue, that is, we get
inspiration from the Shannon entropy concept in order to describe a
different theoretical procedure, hoping that it will be useful to the
area of information theory, in particular, to the study of quantum
information related to the scattering in quantum graphs
\cite{PRL.79.4794.1997,AoP.274.76.1999,AP.55.527.2006,
Book.2012.Berkolaiko,PRL.110.094101.2013}.
As one knows, an important procedure to describe scattering in graphs
comes from the use of Green's function, as it is described in
Refs. \cite{PR.647.1.2016,PRA.98.062107.2018}.
The procedure is well-established but the outcome is somehow involved
since it depends on the energy of the incoming wave, even when one deals
with very simple graphs.

\begin{figure}[t]
  \centering
  \includegraphics[width=\columnwidth]{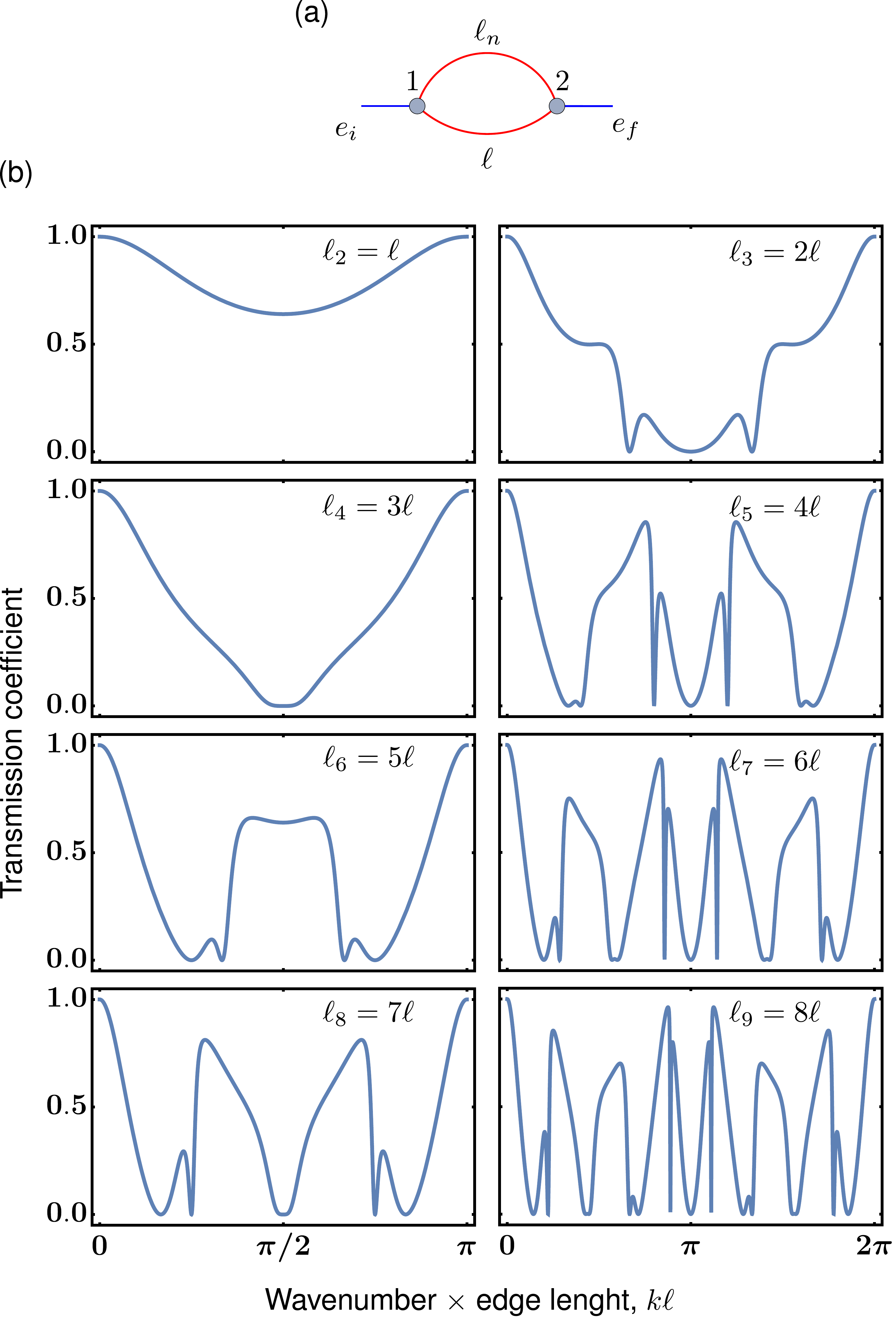}
  \caption{(Color online)
    (a) Quantum graph with two vertices of degree $3$, with
    $\ell_n=(n-1)\ell$.
    (b) Transmission coefficients for $n=2,\ldots,9$.
    For odd $n$ the transmission coefficients have period $\pi/\ell$,
    and for even $n$ the period is $2\pi/\ell$.
  }
  \label{fig:fig1}
\end{figure}

To illustrate this point of view, let us think of the graph that appears
in Fig. \ref{fig:fig1}(a), with $\ell_n=(n-1)\ell$ and $n=2,3,4,...$.
The transmission coefficients are displayed in Fig. \ref{fig:fig1}(b),
in the cases $n=2,\ldots,9$, with $k$ being the wave number associated
with the energy $E=\hbar^2 k^2/2m$ of the incoming wave;
see Sec. \ref{sec:sqg} for more information on how to calculate the
scattering amplitudes.
The results are  analytical and periodic, but more and more complicated,
as we increase $n$ to higher and higher values, even though we keep the
relation $\ell_n/\ell$ an integer.
For this reason, in this work we describe a way to associate a given
graph with a scattering entropy.
Here we focus on the scattering on quantum graphs, but we think the idea
is more general and can be used in other contexts.
To reach this goal, in the Sec. \ref{sec:sqg} we first review the
calculation of the scattering amplitudes for quantum graphs.
We then move on and in Sec. \ref{sec:ase} we introduce and describe how
to calculate the average scattering entropy associated with a given
quantum graph, and in Sec. \ref{sec:examples} we illustrate the general
results with several examples involving different types of quantum
graphs of general interest.
In Sec. \ref{sec:other}, we study several specific issues that appear
very naturally from the results obtained in Sec. \ref{sec:examples}.
We then conclude the work in Sec. \ref{con}, where we comment on the
main results described in the present work and suggest some distinct
lines of investigation related to them.

\section{Scattering in quantum graphs}
\label{sec:sqg}

\begin{figure}[t]
  \centering
  \includegraphics[width=0.7\columnwidth]{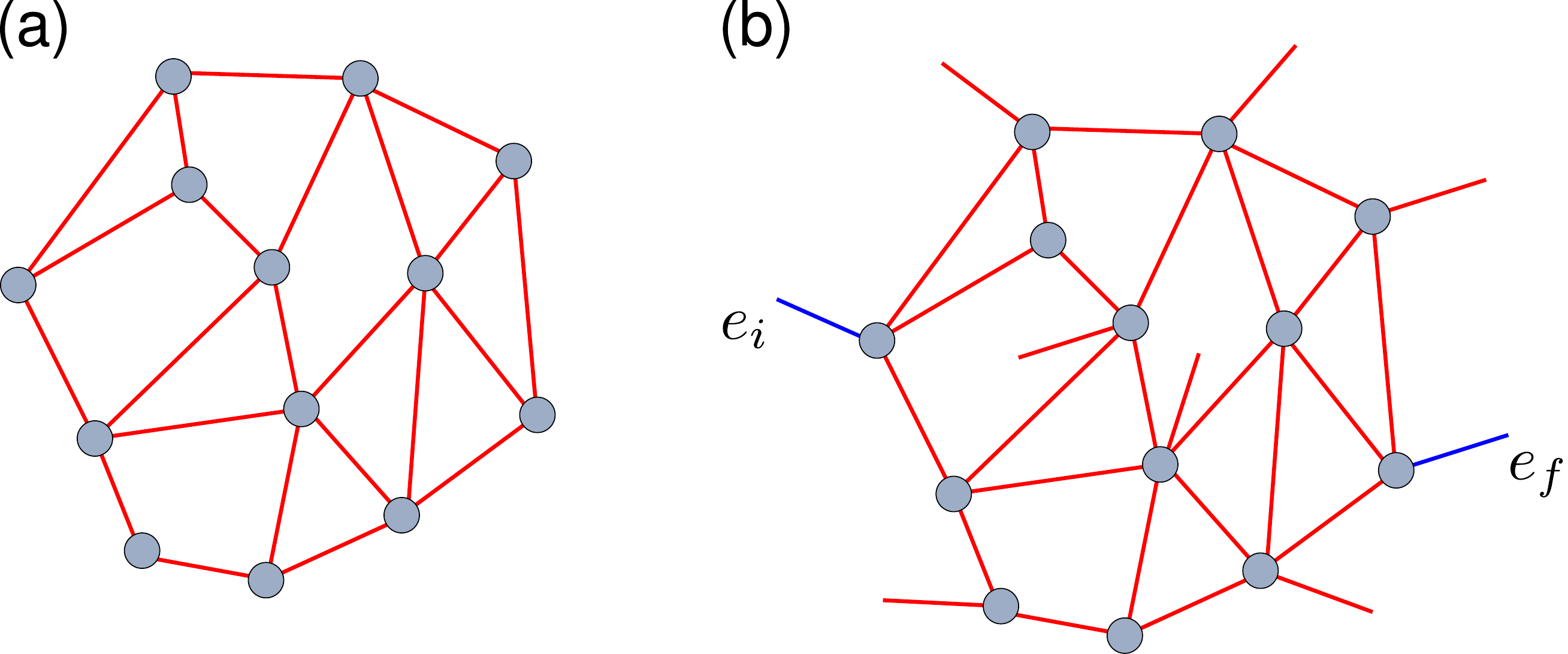}
  \caption{(Color online)
    (a) A quantum graph with $13$ vertices and $23$ edges.
    (b) The same graph but now with $9$ leads added.
    The $2$ blue leads are the entrance and exit leads used in the Green's
    function construction.
  }
  \label{fig:fig2}
\end{figure}

Let us start by considering the scattering in simple quantum graphs.
A quantum graph is a triple $\{\Gamma(V,E), H, \text{BC}\}$,
consisting of a metric graph, $\Gamma(V,E)$; a differential operator,
$H$; and a set of boundary conditions, BC \cite{Book.2012.Berkolaiko}.
A metric graph is a set of $v$ vertices, $V=\{1,\ldots,v\}$, and a set
of $e$ edges, $E=\{e_1,\ldots,e_e\}$, where each edge is a pair of vertices
$e_s=\{i,j\}$, and we assign positive lengths to each edge
$\ell_{e_{s}}\in (0,\infty)$.
Here we consider the free Schr\"odinger operator
$H=-(\hbar^2/2m)d^2/dx^2$ on each edge and the most natural set of
boundary conditions, namely, the Neumann boundary conditions.
The graph topology is totally defined by its adjacency matrix
$A(\Gamma)$ of dimension $v \times v$.
The elements $A_{ij}(\Gamma)$ are $1$ if the vertices $i$ and $j$ are
connected  and $0$ otherwise.
We create an open quantum graph, $\Gamma^{l}$, which is suitable to study
scattering problems, by adding $l$ leads (semi-infinite edges) to its
vertices (see Fig. \ref{fig:fig2}).
The open quantum graph $\Gamma^{l}$ then represents a scattering system
with $l$ scattering channels which is characterized by the energy
dependent global scattering matrix
$\boldsymbol{\sigma}_{\Gamma^{l}}(k)$, where $k=\sqrt{2mE/\hbar^2}$ is
the wave number, and the matrix elements are given by the scattering
amplitudes $\sigma_{\Gamma^{l}}^{(f,i)}(k)$, where $i$ and $f$ are the
entrance and exit scattering channels, respectively.

In this work, we employ the Green's function approach as developed in
Refs. \cite{PRA.98.062107.2018,PR.647.1.2016} to determine the
scattering amplitudes $\sigma_{\Gamma^{l}}^{(f,i)}(k)$.
This technique was used to study narrow peaks of full transmission and
transport in simple quantum graphs
\cite{PRA.100.62117.2019,EPJP.135.451.2020}, which has inspired us to
describe the present study.
The exact scattering Green's function for a quantum
particle of wave number $k$ entering the graph at the vertex $i$ and
exiting the graph at the vertex $f$ is given by
\begin{equation}
  \label{eq:GF}
  G_{\Gamma^{l}}^{(f,i)}(k) = \frac{m}{i \hbar^2 k}
  \left[
  \delta_{fi} e^{i k |x_f-x_i|}+
  \sigma_{\Gamma^{l}}^{(f,i)}(k) e^{i k (|x_{f}|+|x_{i}|)}
  \right],
\end{equation}
where
\begin{equation}
  \label{eq:Srs}
  \sigma_{\Gamma^{l}}^{(f,i)}(k) =
  \delta_{fi}r_{i}
  +
  \sum_{j \in E_{i}}  A_{ij} P_{ij}^{(f)}t_{i}.
\end{equation}
In Eq. \eqref{eq:GF}, $x_i$ and $x_f$ are reference points on the leads
connected to the vertices $i$ and $f$, respectively.
In Eq. \eqref{eq:Srs}, $E_{i}$ is the set of neighbor vertices
connected to $i$, and  $r_{i}$ ($t_{i}$) is the $k$-dependent
reflection (transmission) amplitude at the vertex $i$.
Moreover, $P_{ij}^{(f)}$ is the family of paths between the vertices $i$
and $j$, which are given by
\begin{equation}
  \label{eq:pij}
  P_{ij}^{(f)}
  =
   z_{ij}\delta_{fj} t_{j}
  +
  z_{ij}  P_{ji}^{(f)} r_{j}
  +z_{ij} \sum_{l \in {E_{j}^{i,f}}}  A_{jl} P_{jl}^{(f)} t_{j},
\end{equation}
where $z_{ij}= e^{i k \ell_{s}}$ with $\ell_{s}$ being the length of the
edge $e_{s}=\{i,j\}$ connecting $i$ and $j$, and $E_{j}^{i,f}$ being the
set of neighbors vertices of $j$ but with the vertices $i$ and $f$
excluded.
The family $P_{ji}^{(f)}$ can be obtained from the above equation by
swapping $i \leftrightarrow j$, and the number of family of paths is
always twice the number of edges in the underlying graph.
The family of paths altogether form an inhomogeneous system of equations
that can be written as
\begin{equation}
  \label{eq:system}
  \mathbf{P}_{\Gamma^{l}} = U_{\Gamma^{l}}(k) \mathbf{P}_{\Gamma^{l}}
  + \mathbf{Z}_{\Gamma^{l}},
\end{equation}
where $\mathbf{P}_{\Gamma^{l}}=
(P_{12}^{(f)},P_{21}^{(f)},\ldots,P_{n-1n}^{(f)},P_{nn-1}^{(f)})^{T}$ ,
$\mathbf{Z}_{\Gamma^{l}}=
(\delta_{1f} z_{1f} t_{f},\ldots,\delta_{nf} z_{nf}t_{f})^{T}$,
and $U_{\Gamma^{l}}(k)$ is a unitary quantum evolution map which can be
factored as a product of two unitary matrices \cite{PRA.98.062107.2018}:
\begin{equation}
U_{\Gamma^{l}}(k)=D_{\Gamma^{l}}(k)S_{\Gamma^{l}}(k),
\end{equation}
where $D_{\Gamma^{l}}(k) = \diag(z_{12},z_{12},\ldots,z_{n-1n},z_{n-1,n})$
and $S_{\Gamma^{l}}(k)$ is the scattering matrix of the graph.
Following Eq. \eqref{eq:Srs}, the solution of Eq. \eqref{eq:system}
provides the scattering amplitude $\sigma_{\Gamma^{l}}^{(f,i)}(k)$
\cite{PRA.98.062107.2018} and it involves the secular determinant
\begin{equation}
  \label{eq:secular}
  \zeta_{\Gamma^{l}}(k)=\det\left[\mathbbm{1}-U_{\Gamma^{l}}(k)\right].
\end{equation}
This determinant is very important in quantum graph theory
\cite{PRL.79.4794.1997,AoP.274.76.1999,AP.55.527.2006} and is used to
study bound states and resonances which can be obtained by considering
the roots of $\zeta_{\Gamma^{l}}(k)=0$.

The individual quantum scattering amplitudes $r_i$ and $t_i$ at each
vertex are determined by the boundary conditions and are, in general,
$k$-dependent.
As stated above, here we use Neumann boundary conditions,
which lead to $k$-independent quantum amplitudes \cite{AP.55.527.2006},
namely,
\begin{equation}
  \label{eq:scatt_amp}
  r_{i} = \frac{2}{d_{i}} -1, \qquad
  t_{i} = \frac{2}{d_i},
\end{equation}
where $d_i\geq 2$ is the degree of the vertex $i$ (the total number of
edges and/or leads attached to it).
In this case, the vertex is usually called a Neumann vertex.
It is also
important to observe that even though the above individual quantum
amplitudes are $k$-independent the scattering amplitudes
$\sigma_{\Gamma^{l}}^{(f,i)}(k)$ are $k$-dependent.
For the case where the vertex degree is $1$, i.e., when the vertex is a
dead end, the Neumann and Dirichlet boundary conditions lead to $r_i=1$
and $-1$, respectively
\cite{PR.647.1.2016}.

\section{Average scattering entropy of quantum graphs}
\label{sec:ase}

Now consider again the scattering quantum graph $\Gamma^{l}$ described
in the previous section.
By fixing the entrance channel, say $i$, this scattering system is
characterized by $l$ quantum amplitudes, which are obtained
\textit{analytically} from the Green's function in Eq. \eqref{eq:GF},
and defines a set of $l$ quantum probabilities as
\begin{equation}
  p_{\sigma_{\Gamma^{l}}}^{(j)}(k)=|\sigma_{\Gamma^{l}}^{(j,i)}(k)|^{2},
\end{equation}
which are the probabilities for a particle entering the graph, with wave
number $k$, by the fixed vertex $i$ and exiting the graph by the vertex
$j$ (including the vertex $i$ itself),
and fulfills the relation
\begin{equation}
  \sum_{j=1}^{l}p_{\sigma_{\Gamma^{l}}}^{(j)}(k)=1.
\end{equation}
Then, when a scattering process occurs in a graph, it responds with
$l$ distinct probabilities (the scattering probabilities), in a way
similar to a discrete random variable with $l$ possible outcomes.
So, in direct analogy with the Shannon entropy of a discrete random
variable, here we define the entropy
\begin{equation}
  \label{eq:S_perk}
  H_{\sigma_{\Gamma^{l}}} (k)=
  -\sum_{j=1}^{l} p_{\sigma_{\Gamma^{l}}}^{(j)}(k)
  \log_2 p_{\sigma_{\Gamma^{l}}}^{(j)}(k).
\end{equation}
The above quantity encodes the informational content as a function of
$k$ of the scattering process in a graph.
Note that when all the $l$ scattering channels have the same
probability, $p_{\sigma_{\Gamma^{l}}}^{(j)}(k)=1/l$,
$H_{\sigma_{\Gamma^{l}}} (k)$ assumes its
maximum value $\log_2l$, and the minimum value 0 occurs when all the
quantum probabilities are $0$ but one is equal to 1, for example, in the
case of a full reflection or a full transmission.
The case where $l=2$ is equivalent to a Bernoulli random variable where
$p_{\sigma_{\Gamma^2}}^{(f)}=|\sigma_{\Gamma^2}^{(f,i)}|^{2}$ and
$p_{\sigma_{\Gamma^2}}^{(i)}= |\sigma_{\Gamma^2}^{(i,i)}|^{2}=
1-p_{\sigma_{\Gamma^2}}^{(f)}=1-|\sigma_{\Gamma^2}^{(f,i)}|^{2}$.
The entropy in this case is then
\begin{align}
  H_{\sigma_{\Gamma^{2}}}(k) = {}
  & -|\sigma_{\Gamma^{2}}^{(f,i)}|^{2} \log_2|\sigma_{\Gamma^{2}}^{(f,i)}|^{2}
    \nonumber \\
  &-(1-|\sigma_{\Gamma^{2}}^{(f,i)}|^{2})
  \log_2(1- |\sigma_{\Gamma^{2}}^{(f,i)}|^{2}).
\end{align}

Therefore, for graphs where the scattering amplitudes have a period $K$,
as is the case in this work, we define the {\it average scattering entropy} in the form
\begin{equation}
  \label{eq:AS}
  \bar{H}(\sigma_{\Gamma^{l}})=\frac{1}{K}
  \int_0^K H_{\sigma_{\Gamma^{l}}} (k)\, dk,
\end{equation}
which translates all the complicated behavior of the $k$-dependent
quantum probabilities along the period $K$ to a global information which
is independent of $k$.
In what follows we shall consider equilateral quantum graphs in
which $z_{ij}=z=e^{i k \ell}$, and we shall study how
$\bar{H}(\sigma_{\Gamma^{l}})$ behaves in several distinct situations.
As far as we know, the above procedure which deals with scattering in
quantum graphs is original and may also work in other situations of
current interest, in particular, in the case where quantum scattering is
an important outcome.

\section{Examples}
\label{sec:examples}

\begin{figure}[t]
  \centering
  \includegraphics[width=\columnwidth]{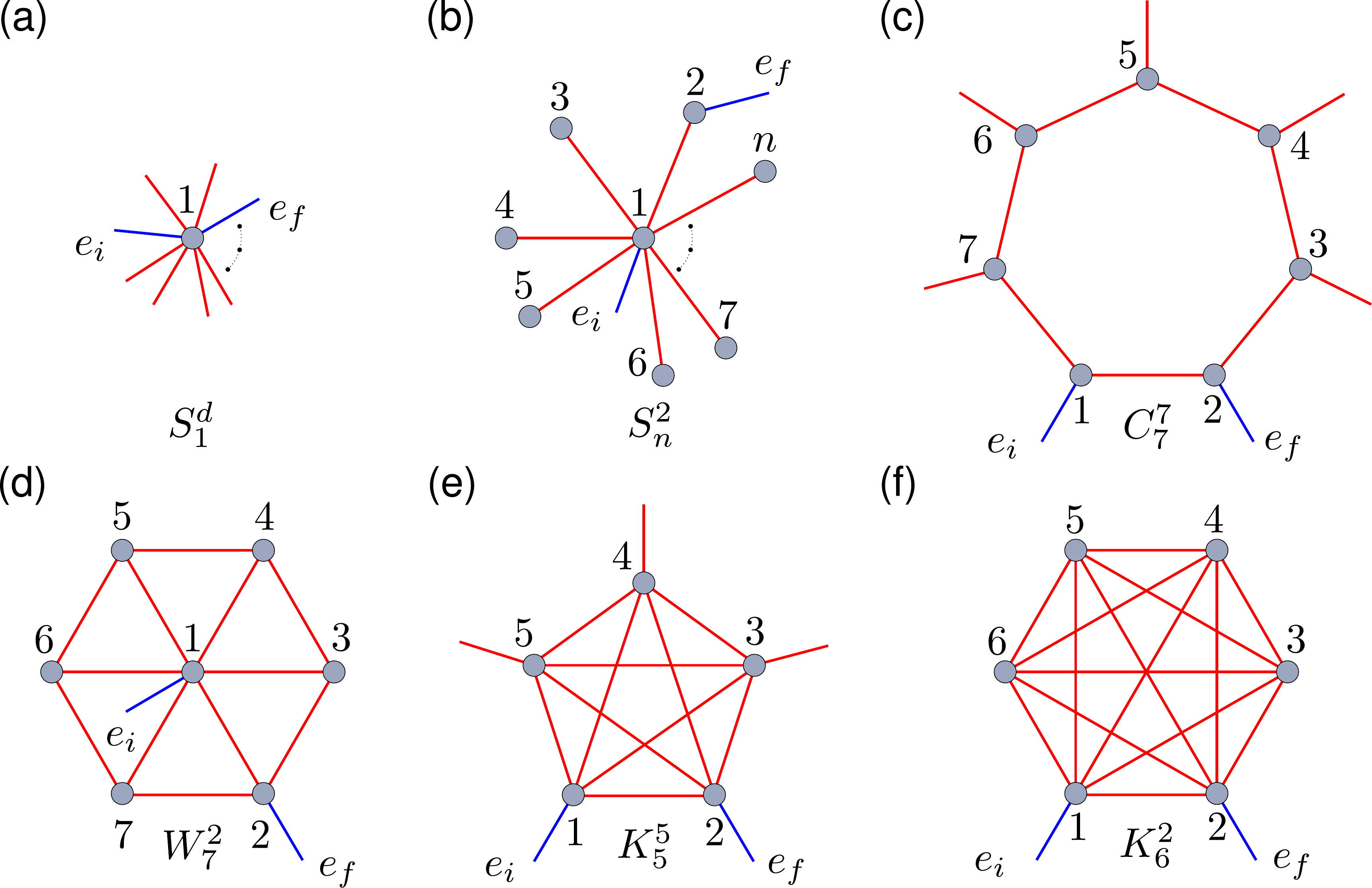}
  \caption{(Color online)
    Examples of graphs considered in this work:
    (a) a single vertex of degree $d$, $S_1^{d}$;
    (b) a star graph with $n$ vertices and two leads added, $S_n^{2}$;
    (c) a cycle graph with seven vertices and with one lead added to
    each vertex, $C_7^{7}$,
    (d) the wheel graph with seven vertices and with two leads added,
    $W_7^{2}$;
    (e) a complete graph with five vertices  with one lead attached to
    each vertex, $K_5^{5}$;
    and (f) a complete graph with six vertices with two leads added,
    $K_6^{2}$.
  }
  \label{fig:fig3}
\end{figure}

Let us now calculate the average scattering entropy for some types of
graphs \cite{Book.2010.Diestel} where we add only two leads to the
graph, as we illustrate in Fig. \ref{fig:fig3}(b),
and when we add one lead to each vertex of the graph, as schematically
shown in Fig. \ref{fig:fig3}(c).

\subsection{Graphs with a single vertex}

\begin{figure}[t!]
  \centering
  \includegraphics[width=\columnwidth]{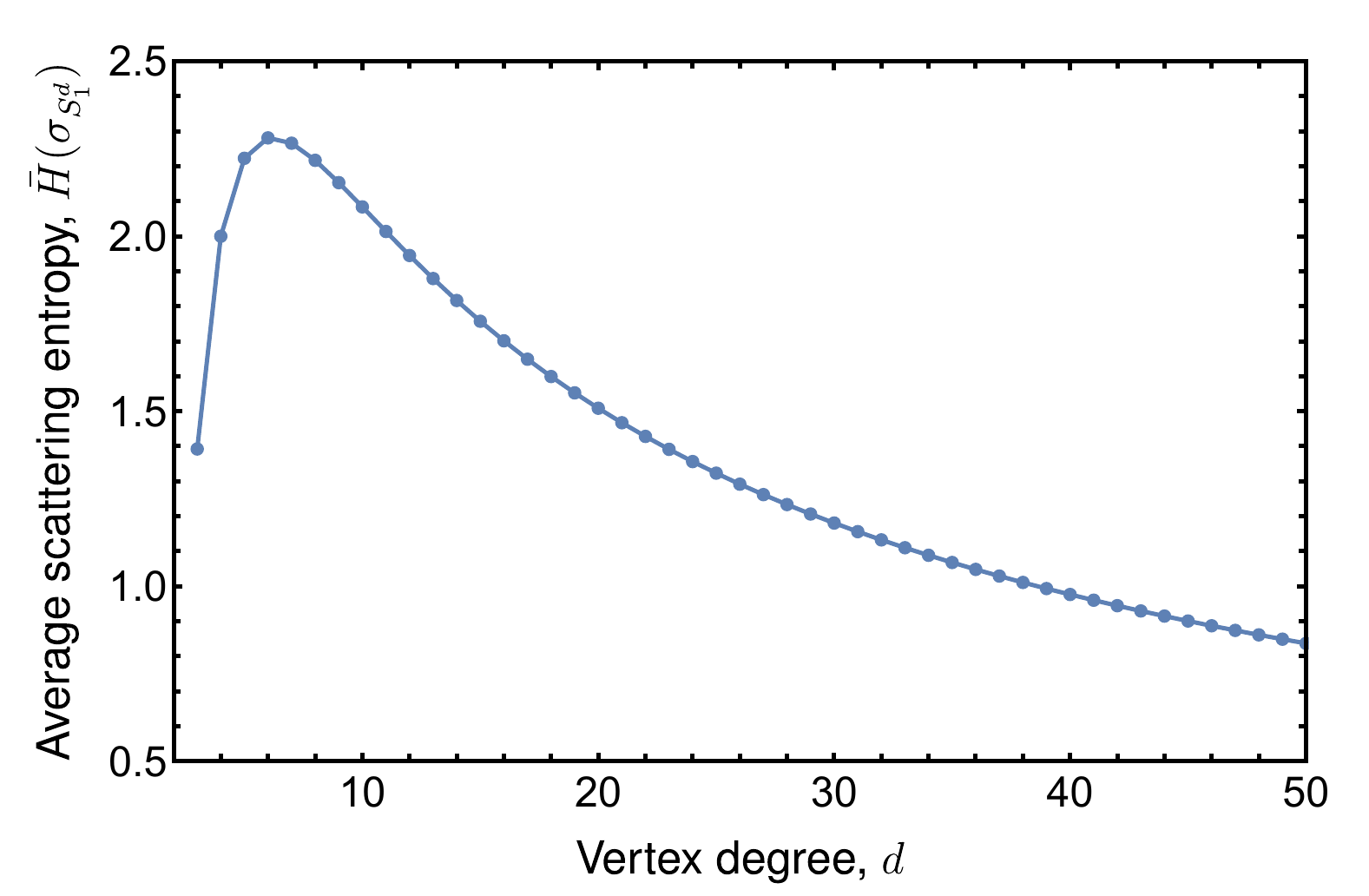}
  \caption{(Color online)
    Behavior of $\bar{H}(\sigma_{S_1^{d}})$ for a graph with a single
    Neumann vertex as a function of its degree $d$.
  }
  \label{fig:fig4}
\end{figure}

Let us start by the simplest case of a quantum graph with a single
vertex which is denoted by $S_1$.
Thus, by adding just one lead we turn it into $S_1^{1}$.
Being a dead end vertex, it is completely opaque and an incoming wave
would be fully reflected, leading to $\bar{H}(\sigma_{S_1^{1}})=0$.
Next, consider a single vertex with the Neumann boundary condition and
two leads, $S_{1}^{2}$.
In this case, the graph is transparent and an incoming wave would be
fully transmitted and $\bar{H}(\sigma_{S_1^{2}})=0$.
These two simple situations represent the two extremes cases of the
biased coins, and nontrivial possibilities appear when we consider a
vertex with a degree larger than $2$.
Then, for a single Neumann vertex of degree $d\geq 3$ [see
Fig. \ref{fig:fig3}(a)] and considering the scattering amplitudes in
Eq. \eqref{eq:scatt_amp}, we arrive at the following analytical result
for the average scattering entropy
\begin{equation}
  \bar{H}(\sigma_{S_1^{d}})=
  \frac{4(d-1)}{d^2}\log_2\frac{d^2}{4}+
  \frac{(d-2)^2}{d^2}\log_2\frac{d^2}{(d-2)^2}.
\end{equation}
In Fig. \ref{fig:fig4} we show the behavior of
$\bar{H}(\sigma_{S_1^{d}})$ as a function of the vertex degree $d$.
We can observe that $\bar{H}(\sigma_{S_1^{d}})$ has its maximum for
$d=6$, and by increasing $d$, $\bar{H}(\sigma_{S_1^{d}})$
vanishes asymptotically as a consequence of the increasing of the
reflection probability.

\subsection{Star graphs $S_n$}
\label{sec:Sn}

\begin{figure}[t!]
  \centering
  \includegraphics[width=\columnwidth]{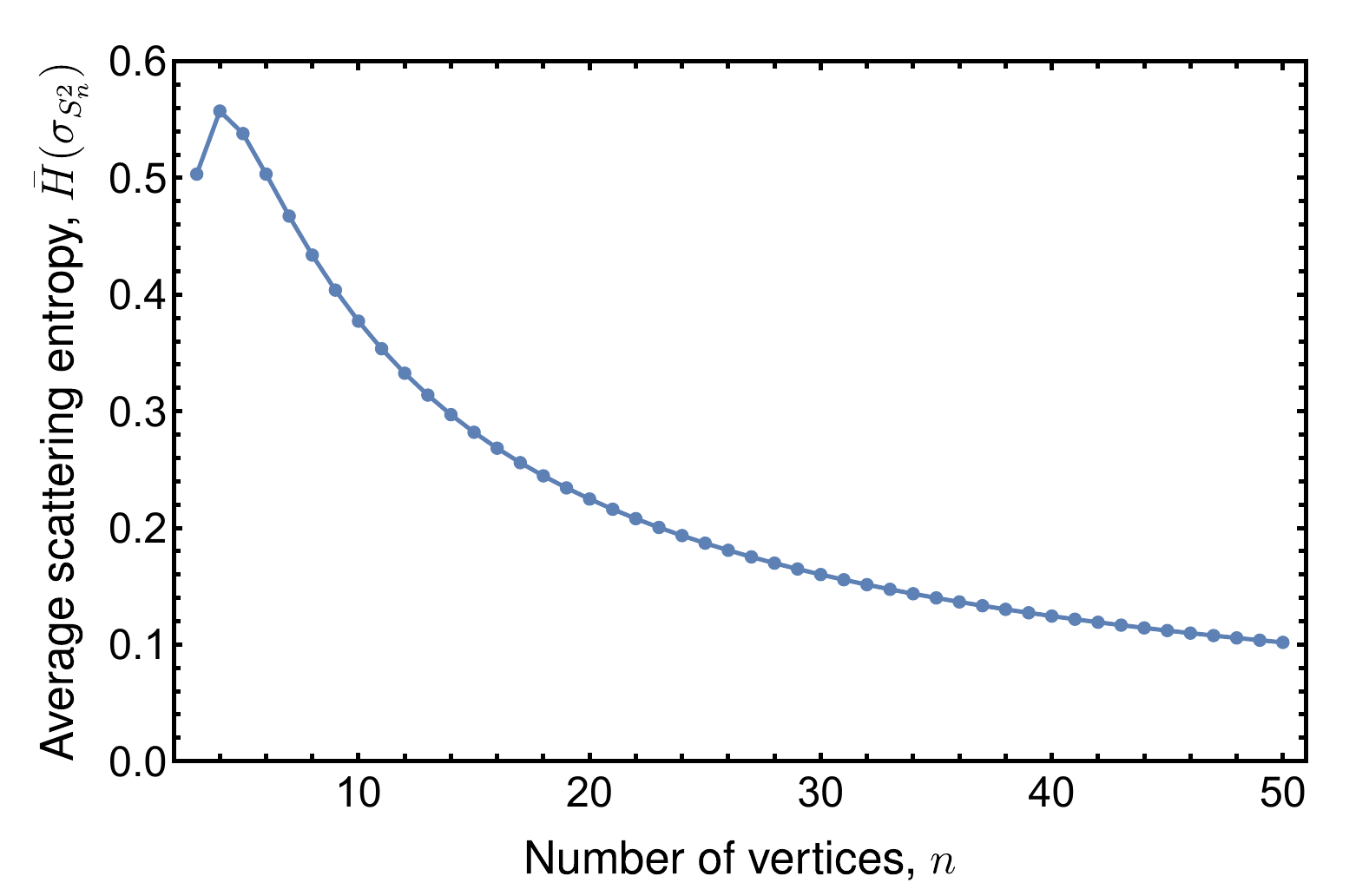}
  \caption{(Color online)
    Behavior of the average entropy $\bar{H}(\sigma_{S_n^{2}})$ for star
    graphs as a function of $n$ for $n=3,...,50$.
  }
  \label{fig:fig5}
\end{figure}

Consider now star graphs $S_n$ with $n$ vertices, graphs where the
central vertex has degree $n-1$ and all others vertices have degree $1$.
We can add one lead to each one of the vertices of the graph, but
as we are employing Neumann vertices, this case should be equivalent
to the previous system of a graph with a single vertex.
Then, we add only two leads, one at the central vertex $1$ and
the other at vertex $2$ as depicted in Fig. \ref{fig:fig3}(b),
leading to a scattering system $S_n^{2}$ with two scattering channels in
which the central vertex has degree $n$.
In this case, using the Green's function approach
\cite{PRA.98.062107.2018} and the Neumann boundary condition, it is
possible to determine analytically the transmission amplitude as a
function of $n$ ($n \geq 3$), which is given by
\begin{equation}
  \label{eq:sigmaSn2}
  \sigma_{S_n^{2}}^{(j,1)}(k)=
  \frac{2z(1+z^2)^{n-2}}{n \zeta_{S_n^{2}}},
\end{equation}
where $z=e^{i k \ell}$ and
\begin{equation}
  \zeta_{S_n^{2}}
  = \det[\mathbbm{1}-U_{S_n^{2}}]
  = \frac{[n-(n-4)z^2](1+z^2)^{n-3}}{n},
  \end{equation}
is the secular determinant.
It is important to comment that in comparison with the previous result
for the quantum amplitudes for $S_1^{d}$ quantum graphs, which are
$k$-independent, the scattering amplitude in Eq. \eqref{eq:sigmaSn2} is
now $k$-dependent.
So, using Eq. \eqref{eq:sigmaSn2} we can calculate the average
scattering entropy and the results are shown in Fig. \ref{fig:fig5} as a
function  of $n$.
We observe that it also has a maximum, which occurs here for $n=4$.
By increasing $n$, the transmission probability in this case diminishes
and the reflection tends to $1$, making the average scattering entropy
vanish for very large values of $n$.
We have reviewed the calculation using Dirichlet boundary conditions for
the vertices of degree $1$, and the results remained the same.
We then conclude that, for the graphs $S_n^2$, the average scattering
entropy does not depend on the boundary conditions at the vertices of
degree $1$ being Neumann or Dirichlet.

\subsection{Cycle graphs $C_n$}
\label{sec:Cn}

\begin{figure}[t!]
  \centering
  \includegraphics[width=\columnwidth]{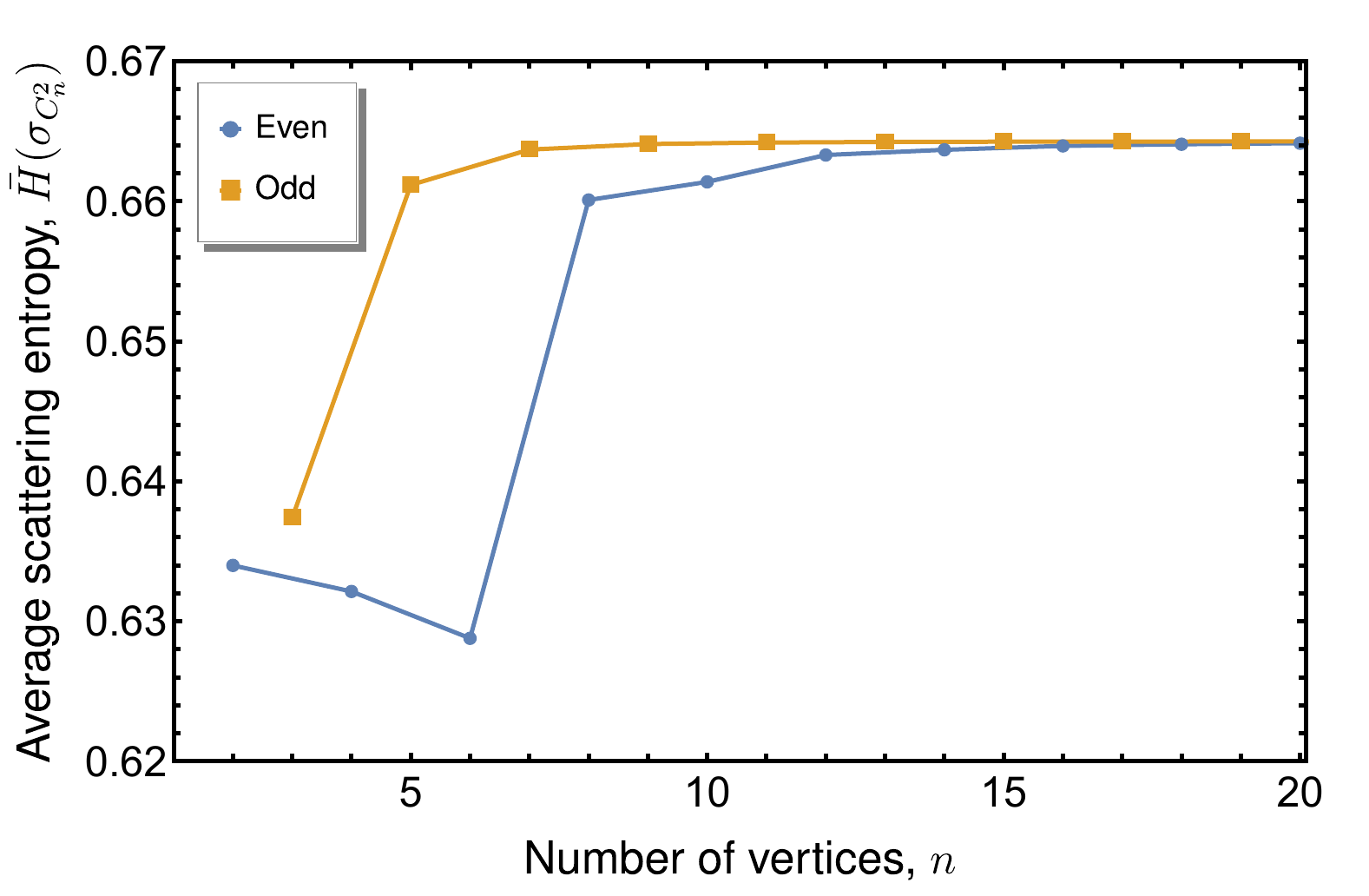}
  \caption{(Color online)
    The average entropy for cycle graphs $C_n^{2}$ as a function of $n$.
    Blue circles are for $n$ even, and orange squares are for $n$ odd.}
  \label{fig:fig6}
\end{figure}

A cycle graph $C_n$ consists of a set of $n$ vertices connected in a
closed chain where each vertex has degree $2$.
Here we study two possible configurations.
In the first configuration we add one lead to vertex $1$ and one to
vertex 2, leading to $C_n^{2}$.
In this case, vertices $1$ and $2$ have degree $3$ and all the other
vertices have degree $2$.
As we are using Neumann vertices, the vertices of degree $2$ are ideal,
being totally transparent.
So, increasing the number of vertices just amounts to increasing the
length of one edge, connecting the entrance and exit vertices
(vertex $1$ and $2$, respectively).
Consequently, this system is equivalent to the cycle graph with two
edges of lengths $\ell$ and $(n-1)\ell$ and two leads discussed in
the Introduction [see Fig. \ref{fig:fig1}(a)], and it is similar to the
ring graph considered in Ref. \cite{APPA.124.1087.2013}, which was used
to study narrow resonances.
However, these narrow resonances are absent in our graphs because here
we employ edges with commensurable lengths.
Thus, with all this in mind, we can calculate analytically the
transmission amplitude which is given by
\begin{equation}
  \label{eq:sigma_Cn2}
  \sigma_{C_{n}^{2}}^{(2,1)}(k) =
  \frac{4z(z^{n}-1)(z^n+z^2)}{9z^2-z^4-z^{2n}-8z^{n+2}+z^{2n+2}},
\end{equation}
and the scattering entropy per $k$ which is given by
\begin{align}
  H_{C_n^{2}}(k) = {}
  & -|\sigma_{C_{n}^{2}}^{(2,1)}|^{2} \log_2|\sigma_{C_{n}^{2}}^{(2,1)}|^{2}
    \nonumber \\
  &-(1-|\sigma_{C_{n}^{2}}^{(2,1)}|^{2})
  \log_2(1- |\sigma_{C_{n}^{2}}^{(2,1)}|^{2}).
\end{align}
The average scattering entropy can be then numerically calculated and
the results are shown in Fig. \ref{fig:fig6}.
We can see  that $\bar{H}(\sigma_{C_n^{2}})$ depends on the parity of
$n$, and for $n$ even, for the first three values $n=2,4$ and $6$, the
average scattering entropy decreases.
This behavior can be understood by looking at the transmission
probabilities in Fig. \ref{fig:fig1}(b).
Comparing the plots on the left, from the top to bottom, we can observe
a change in the shape of the transmission probability for $n=8$ when
compared with $n=2,4$ and $6$.
This change reflects on the behavior of the average scattering entropy.
On the other hand, for $n$ odd or even larger than 6, the average
scattering entropy increases and seems to converge to a constant value.

\begin{figure}[t!]
  \centering
  \includegraphics[width=\columnwidth]{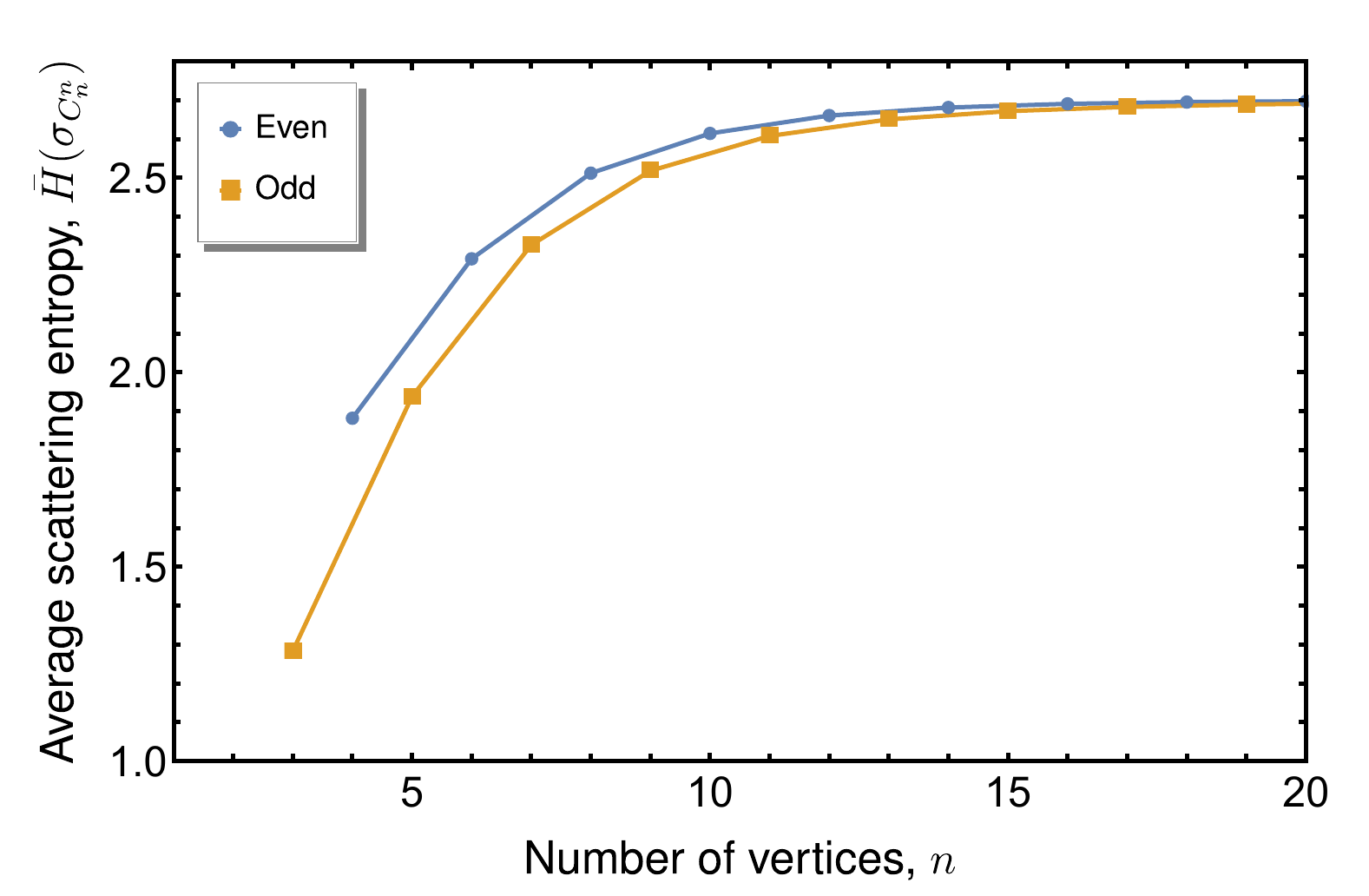}
  \caption{(Color online)
    Behavior of the average scattering entropy for cycle graphs
    $C_n^{n}$ as a function of $n$.
    Blue circles are for $n$ even, and orange squares are for $n$ odd.
  }
  \label{fig:fig7}
\end{figure}

In the second situation we have to add one lead to each vertex, leading
to a graph where all the vertices have degree $3$ [see
Fig. \ref{fig:fig3}(c), for an illustration with seven vertices and
seven leads], and it is now necessary to calculate one reflection and
$n-1$ transmissions amplitudes.
These amplitudes are calculated from the Green's function and their
expressions are quite lengthy to be shown here.
In Fig. \ref{fig:fig7} we show the average scattering entropy as a
function of $n$, and we can observe that it depends on the parity of
$n$ and increases as a function of $n$, reaching a plateau.
As in the previous situation with just two leads, it seems to tend again
to a constant value.
This result is quite interesting because by increasing the value of $n$
we are increasing the number of scattering channels, something that, in
theory, could increase the information content of the system.

\subsection{Wheel graphs $W_n$}
\label{sec:Wn}

\begin{figure}[t!]
  \centering
  \includegraphics[width=\columnwidth]{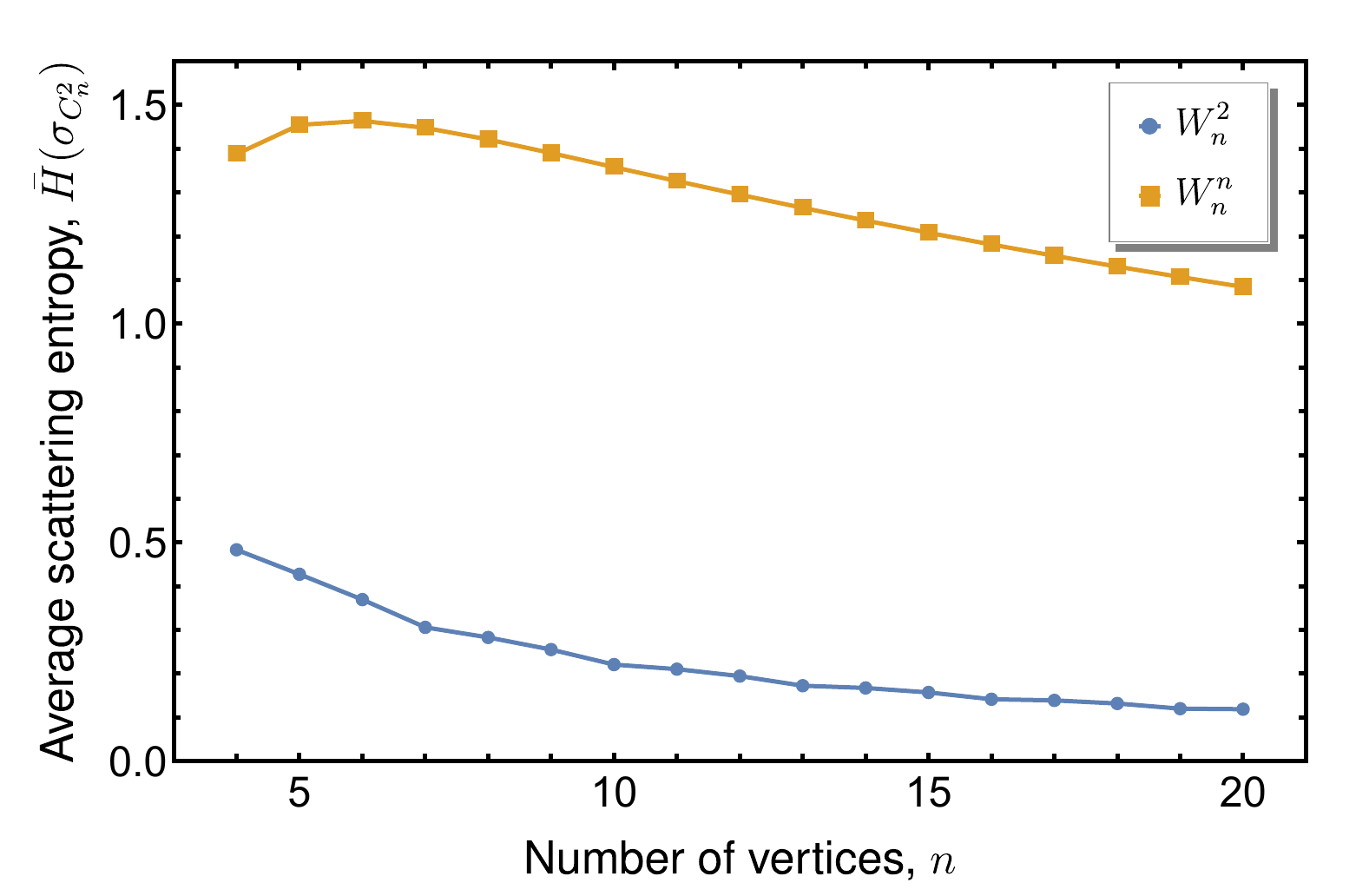}
  \caption{(Color online)
    Behavior of the average scattering entropy for wheel graphs with two
    leads, $W_n^{2}$ (blue circles), and with $n$ leads, $W_n^{n}$ (orange
    squares) as a function of $n$.
  }
  \label{fig:fig8}
\end{figure}

A wheel graph $W_n$ with $n$ vertices is a graph formed by connecting a
single central vertex to $n-1$ peripheral vertices in a circle.
In the first situation we add one lead to the central vertex and one
lead to one of the peripheral vertices leading to $W_n^{2}$ and in the
second situation we add one lead to each vertex leading to $W_n^{n}$.
The expressions for the quantum amplitude are quite complicated
to be shown here and the average scattering entropy is shown in
Fig. \ref{fig:fig8}.
For $W_n^{2}$ its maximum occurs for $n=3$ and decreases to zero for very
large values of $n$.
On the other hand, for $W_n^{n}$ the maximum value occurs for $n=6$, and
it then slowly decreases as $n$ increases in a similar way as for the
$S_1^{d}$ and $S_n^{n}$ graphs.
In all the cases, the increasing of the number of vertices causes the
increasing of the reflection coefficient, which for very large $n$
approaches $1$, inducing the vanishing of the average scattering
entropy.

\subsection{Complete graphs $K_n$}
\label{sec:Kn}

\begin{figure}[t!]
  \centering
  \includegraphics[width=\columnwidth]{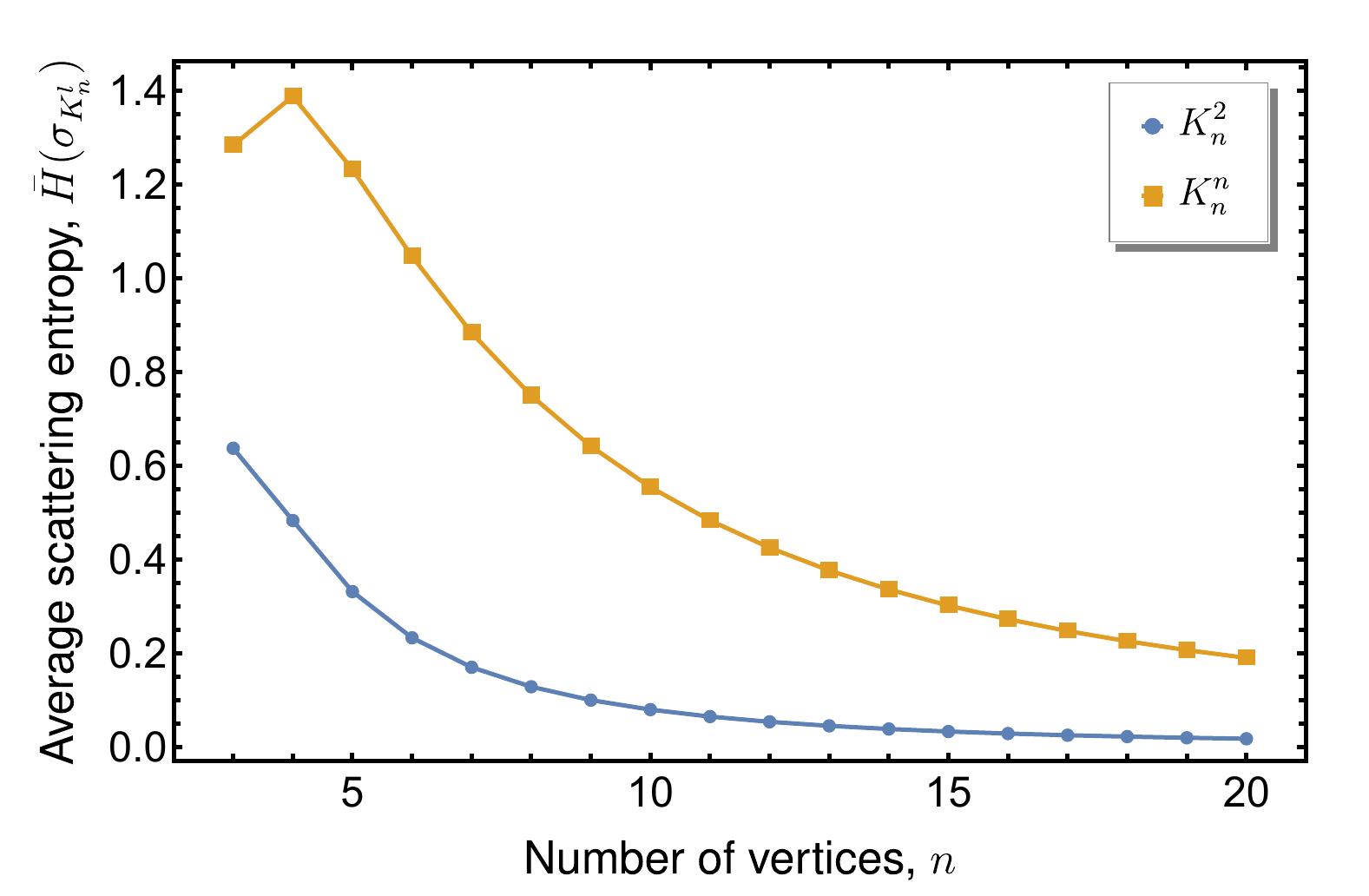}
  \caption{(Color online)
    Behavior of the average scattering entropy for complete graphs with
    two leads, $K_n^{2}$ (blue circles), and with $n$ leads, $K_n^{n}$
    (orange squares), as a function of $n$.
  }
  \label{fig:fig9}
\end{figure}

Now consider complete graphs $K_n$ with $n$ vertices, i.e., graphs in
which every vertex connects to all other vertices via a single edge.
Consider first the graph $K_n$ with one lead attached to  vertex 1
and one lead attached to vertex 2, leading to $K_n^{2}$
[see Fig. \ref{fig:fig3}(f)].
The average scattering entropies for these graphs are shown in
Fig. \ref{fig:fig9} (blue circles) and we can observe that
$\bar{H}(\sigma_{K_n^{2}})$ decreases monotonically in a way similar
to that of $\bar{H}(\sigma_{W_n^{2}})$.
In the case where we attach one lead to each vertex of $K_n$, leading to
$K_n^{n}$ [see Fig. \ref{fig:fig3}(e)], all the vertices have degree
$n+1$.
Although these are the most complex graphs to calculate the quantum
amplitudes, since each vertex is connected to each other, all the
vertices are equivalent.
Consequently, there is one reflection amplitude and $n-1$
identical transmission amplitudes, which leads to  the following
expression for the scattering entropy per $k$:
\begin{align}
  H_{K_n^{n}}(k) = {}
  & -|\sigma_{K_n^{n}}^{(1,1)}|^{2}
    \log_2| \sigma_{K_n^{n}}^{(1,1)}|^{2}\nonumber\\
  & -(1- |\sigma_{K_n^{n}}^{(1,1)}|^{2})
    \log_2\frac{1- |\sigma_{K_n^{n}}^{(1,1)}|^{2}}{n-1},
\end{align}
where $\sigma_{K_n^{n}}^{(1,1)}$ is the global reflection amplitude
at vertex $1$.
The resulting average scattering entropy is shown in Fig. \ref{fig:fig9}
(orange squares); it has a maximum for $n=4$ and then decreases when $n$
increases.
For both $K_n^{2}$ and $K_n^{n}$, the average scattering entropies
decrease as $n$ increases, and then they vanish asymptotically, because
the reflection amplitude tends to 1 for higher and higher values of
$n$.

\section{Other results}
\label{sec:other}

\begin{figure}[t!]
  \centering
  \includegraphics[width=0.85\columnwidth]{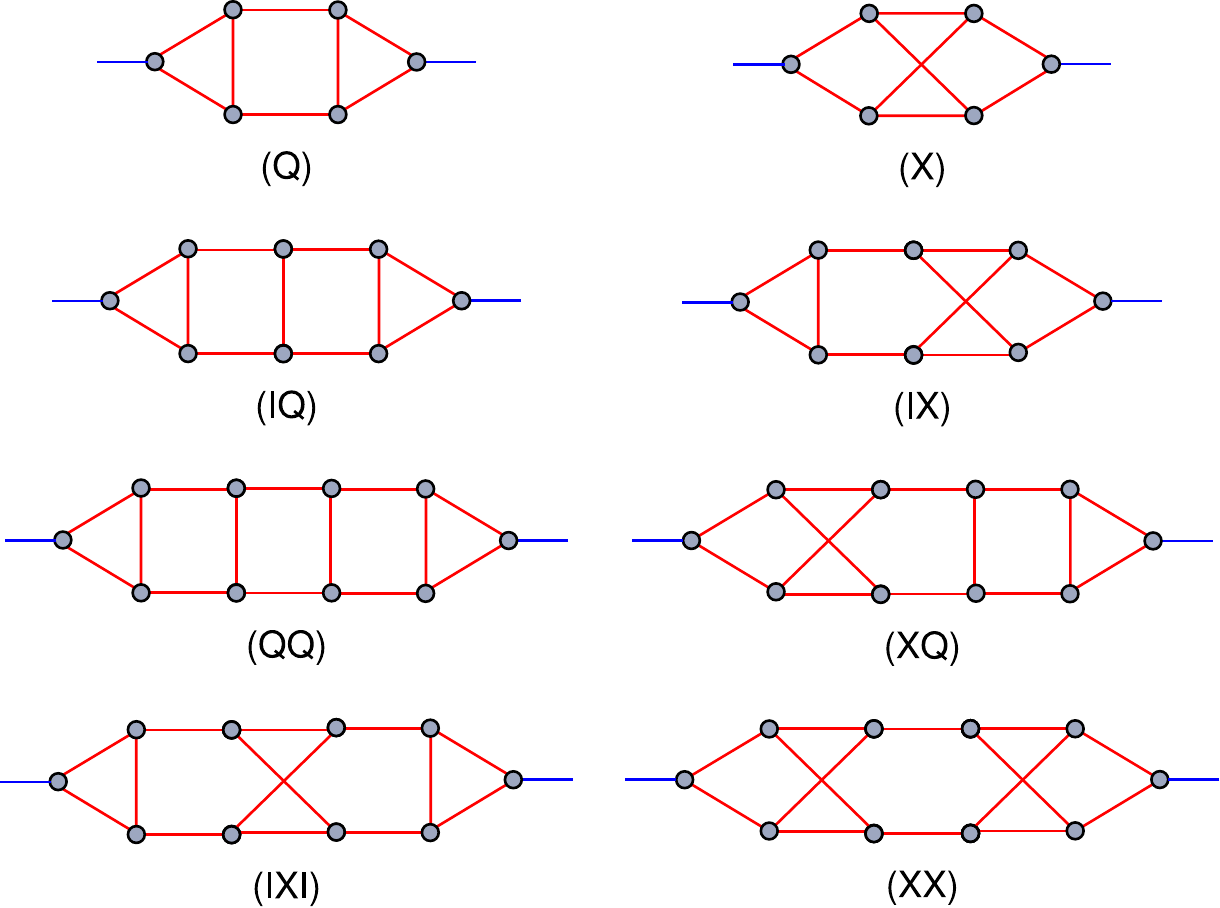}
  \caption{(Color online)
    Graphs Q, X, IQ, IX, QQ, XQ, IXI, and XX, which are considered
    to calculate the average scattering entropy.
  }
  \label{fig:fig10}
\end{figure}

Let us now consider other possibilities.
We first deal with the case where an incoming wave enters the graphs
depicted in Fig. \ref{fig:fig10}.
They are the Q, X, IQ, IX, QQ, XQ, IXI, and XX graphs.
The two first ones, Q and X, were investigated in Ref.
\cite{EPJP.135.451.2020}, and they  have the interesting property of
having the same number of vertices ($6$ vertices), with the very same
degree $3$,
and the same number of edges ($8$ edges), with the very same length
$\ell$, so they only differ from each other topologically.
We use the above methodology to get to the results:
$\bar{H}(\sigma_{\rm Q})=0.634882$ and
$\bar{H}(\sigma_{\rm X}) =0.699852$, so the topology of X has higher
average scattering entropy, and it is favored if one thinks of
describing compositions with higher average scattering entropy.
In the present case the increasing of the average scattering entropy
when one changes from the Q to the X graph is about $10\%$, so it is not
to be discarded.
We now consider two other graphs depicted in Fig. \ref{fig:fig10},
labeled IQ and IX.
They have $8$ vertices and $11$ edges, and the corresponding average
scattering entropies are $\bar{H}(\sigma_{\rm IQ})=0.547333$ and
$\bar{H}(\sigma_{\rm IX})=0.778697$.
In this case the increase in the average scattering entropy is much more
important than in the previous case with the Q and X graphs; so we have
another, more important situation where the two graphs have the same
number of vertices and edges, but the topology is different, and this
gives distinct average scattering entropies.
The other four graphs QQ, XQ, IXI, and XX in Fig. \ref{fig:fig10} have
$10$ vertices of degree $3$, and $14$ edges of length $\ell$.
The corresponding values for the average scattering entropy are
$0.493163$, $0.572996$, $0.582336$, and $0.844156$, respectively, so
they follow a rule which is similar to the rule that appeared before for
the other compositions depicted in Fig. \ref{fig:fig10} and can be used
properly, depending on the necessity of increasing or decreasing the
average scattering entropy.

\begin{figure}[t!]
  \centering
  \includegraphics[width=0.85\columnwidth]{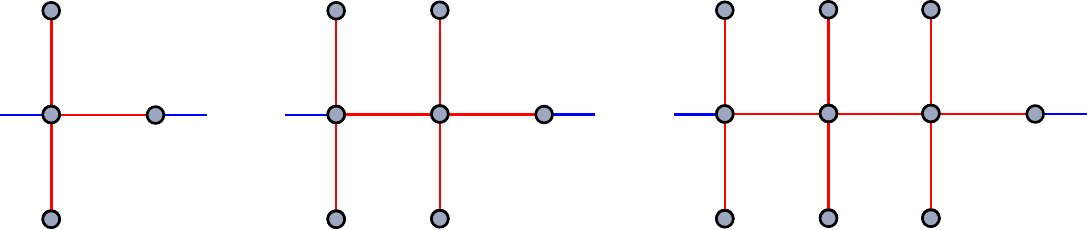}
  \caption{(Color online)
    Graphs ${[S_4^2]}_1$, ${[S_4^2]}_2$, and ${[S_4^2]}_3$, which are
    considered to calculate the corresponding average scattering entropy
    displayed in Fig. \ref{fig:fig12}.
  }
  \label{fig:fig11}
\end{figure}

\begin{figure}[t!]
  \centering
  \includegraphics[width=\columnwidth]{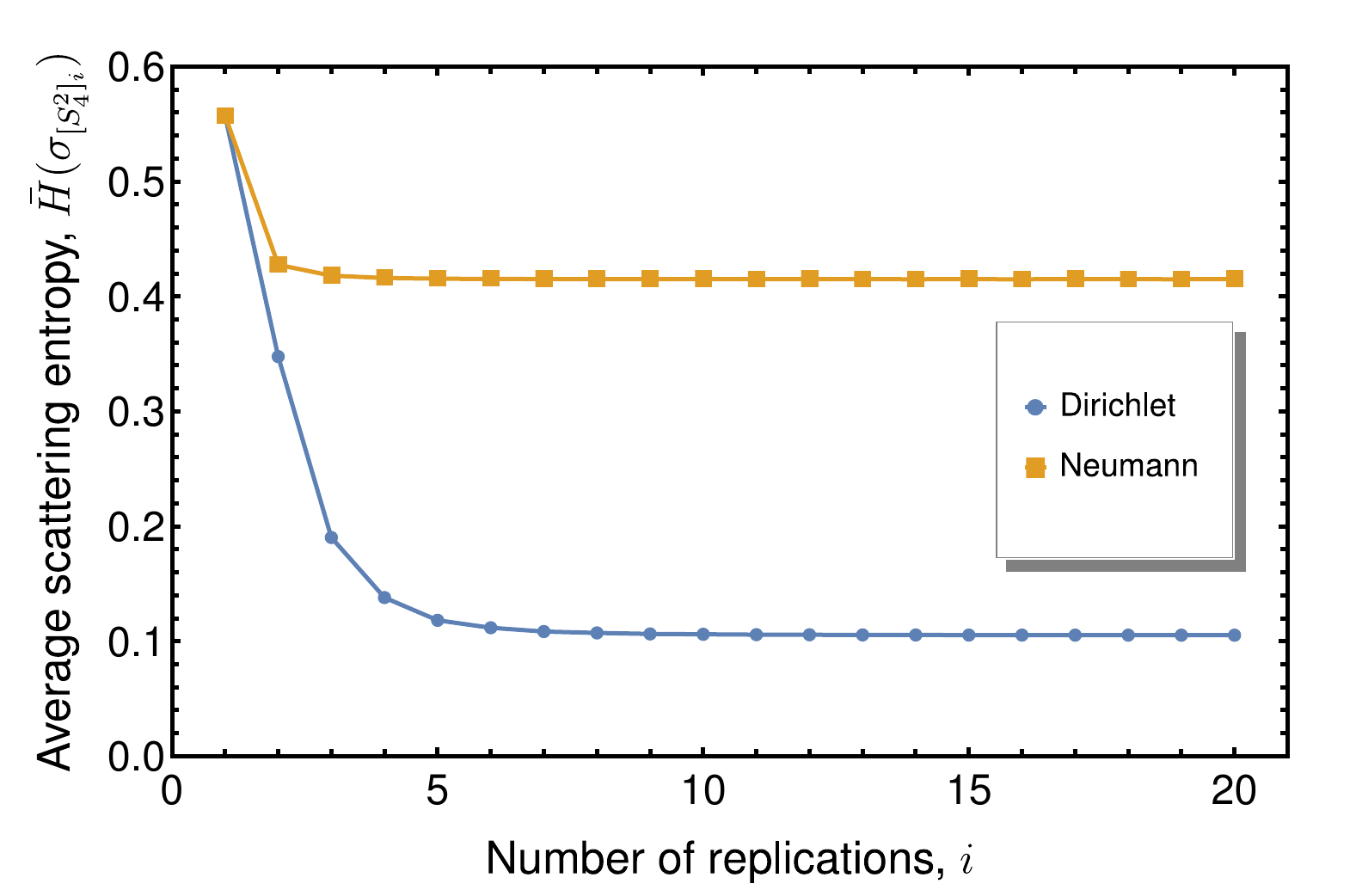}
  \caption{(Color online)
    The average scattering entropy depicted for the ${[S_4^2]}_i$ graphs
    which are illustrated in Fig. \ref{fig:fig11}, for $i=1,2,...,20$.}
  \label{fig:fig12}
\end{figure}

We see from Fig. \ref{fig:fig5} that for the $S^2_n$ type of graph where
the subscript $n$ indicates the number of vertices, and the superscript
$2$  indicates the presence of two leads, one entering at vertex $1$ and
the other leaving at the vertex $2$, as illustrated in
Fig. \ref{fig:fig3}(b), the $S^2_4$ graph is the one with maximum
average scattering entropy, so it can be used for the construction of
other arrangements, if one is thinking of describing situations with
maximum average scattering entropy.
For instance, in Fig. \ref{fig:fig11} we depict some graphs in the
family which we call ${[S_4^2]}_i$, with $i=1,2,3,\cdots$
The arrangements remind us of fishbone structures, with the subscript
$i$ now indicating the number of copies of the elementary structure that
is in the graph.
In Fig. \ref{fig:fig12} we show how the average scattering entropy
varies as we increase $i$.
The results are interesting since they show that the boundary conditions
at the vertices of degree $1$ changes the corresponding quantities: as
we see from Fig. \ref{fig:fig12}, the average scattering entropy is
higher when one uses the Neumann boundary condition, for $i\geq 2$.
We also notice that, as we increase $i$, the numerical values decrease
until reaching an almost constant value as $i$ increases above $5$.
In the case of the Neumann boundary condition, the values of the average
scattering entropy for $i=1$ and $i=2$ are $0.557305$ and $0.427590$,
respectively, with all the other values being essentially the same as in
the case of $i=2$.
This shows that, from the point of view of the average scattering
entropy, the fishbone lattice with several elements is almost
insensitive to the addition or removal of some of its elements,
although it strongly depends on the boundary conditions on the vertices
of degree 1 being Neumann or Dirichlet.

\section{Conclusion}
\label{con}

In this work we have introduced and discussed the entropy associated with
the scattering in simple quantum graphs.
The investigation is inspired by the Shannon entropy concept and by the
quantum scattering in simple quantum graphs.
As we have commented before, the probability that describes the
transmission through a simple quantum graph may be periodic when one
deals with ideal vertices and edges, with the size of the edges
described by integers times a specific length $\ell$.
However, it in general depends on the wave number of the incident wave
and this leads us to reflection and transmission coefficients with
complex shapes even for these simple systems.
In this sense, in order to improve the understanding of the scattering
process that happens in simple graphs, we have proposed a tool to be
associated with the scattering in quantum graphs.
It is the average scattering entropy, that is, the Shannon entropy
described in terms of the probability of the wave being transmitted or
reflected by a quantum graph.

As we have noticed, the several investigations developed in the present
work  have shown that the average scattering entropy works nicely and
can be used as a methodology to add further information about the
issue concerning the scattering in quantum graphs.
In particular, we have obtained the average scattering entropy for
several distinct types of graphs, which are illustrated in
Fig. \ref{fig:fig3}, among them the star $S_n$, the circle $C_n$, the
wheel $W_n$, and the complete $K_n$ graphs.
In this work, we have focused on the scattering in a quantum graph,
which is a triple describing a set of vertices and edges, a differential
operator, and a set of boundary conditions.
Accordingly, the scattering depends on the number and the degree of the
vertices, the number and the length of the edges, the topology, the
scattering channels, and the wave vector $k$ or the energy $E$.
In this sense, although we have observed, for instance, that the average
scattering entropy for cycle graphs saturates and for complete graphs it
quickly vanishes when we increase the number of vertices, we have been
unable to find a direct connection between the average scattering
entropy and the complexity of the quantum graphs investigated in the
present study.
However, we have also considered the graphs depicted in
Fig. \ref{fig:fig10} and shown that they may have very different values
for their entropies, even though they do not differ in the number of
vertices and edges, something that indicates that the average scattering
entropy distinguishes the topologies of the graphs.
This fact motivates us to investigate the connection between
the average scattering entropy introduced in this work and the graph
complexity, an issue which is presently under consideration.

We have also considered the graphs illustrated in Fig. \ref{fig:fig11}
and shown how the corresponding entropies vary as we increase the number
of the elementary structures that compose the graph and as we change the
boundary condition from Neumann to Dirichlet in the vertices of degree
$1$.
This is another result which may find applications in the study of
systems of current physical interest.
The quantum transport in carbon nanotubes that appears in Ref.
\cite{RMP.87.703.2015} is an important possibility, since nanotubes
provide almost one-dimensional systems where electronic disorder can be
reduced to a low level.
Another possibility concerns applications to describe molecules and
polymers, for instance, the quantum transport in single-molecule
junctions, that is, in devices in which a single molecule is
electrically connected by two electrodes; see, e.g., the
recent review \cite{NRP.1.381.2019} and references therein.
Although we have investigated the Shannon entropy associated with the
scattering in quantum graphs, the basic idea presented here
can also be used in other cases involving quantum scattering.
We are now studying other simple graphs where the topology and the
boundary conditions are different, and we are also interested in the
case where the relation between the lengths of edges is not an integer
anymore.
This is a much harder situation and the scattering amplitudes may not be
periodic anymore, but we can still define the entropy as a limiting
process.
Another specific problem concerns the scattering in more general
networks, in particular in a network where the edges may still have the
same length, but be distributed randomly in the network.

Another line of investigation concerns the very recent work of Melnikov
\cite{arXiv:2012.13612}, in which the author studied issues related to
the problem of signal transmission in quantum mechanics in terms of
topological theories, using the analogy between quantum amplitudes and
knot diagrams.
The results suggest that the variation of topology of the topological
device is somehow similar to the transmission of information on quantum
graphs.
In this sense, we think that it is possible to further explore this
possibility in connection with the average scattering entropy which we
described in the present work.
We hope to report on some of the above issues in the near future.

\section*{Acknowledgments}
This work was partially supported by the Brazilian agencies Conselho
Nacional de Desenvolvimento Cient\'ifico e Te\-cnol\'ogico (CNPq),
Funda\c{c}\~{a}o Arauc\'{a}ria (FAPPR, Grant No. 09/2016), Instituto
Nacional de Ci\^{e}ncia e Tecnologia de Informa\c{c}\~{a}o Qu\^{a}ntica
(INCT-IQ), and Para\'iba State Research Foundation (FAPESQ-PB, Grant
No. 0015/2019).
It was also financed by the Co\-or\-dena\c{c}\~{a}o de
Aperfei\c{c}oamento de Pessoal de N\'{i}vel Superior
(CAPES, Finance Code 001).
F.M.A. and D.B. also acknowledge CNPq Grants No. 313274/2017-7 (F.M.A.),
No. 434134/2018-0 (F.M.A.), No. 314594/2020-5 (F.M.A.),
No. 303469/2019-6 (D.B.) and No. 404913/2018-0 (D.B.).

\bibliographystyle{apsrev4-2}
%

\end{document}